\documentstyle[12pt]{article}
\newcommand{\doublespace}{\renewcommand{\baselinestretch}{1.75}
   \Large\normalsize}

\renewcommand{\ref}[1]{\raisebox{.6ex}{[#1]}}

\newcommand{\be}{\begin{equation}}
\newcommand{\ee}{\end{equation}}

\begin{document}
\begin{flushright}
CERN-TH.7385/94
\end{flushright}

 \doublespace

\begin{center}
{\bf   EFFECTIVE LAGRANGIANS FOR BCS SUPERCONDUCTORS AT T=0 }\\
\vspace*{0.5cm}

{\bf Ian J.R. Aitchison$^{1,2,\dag}$,
 Ping Ao$^{2,\#}$, David J. Thouless$^2$, and X.-M. Zhu$^{2,\#}$} \\
\vspace*{0.3cm}
$^{1}$TH. Division, CERN, CH-1211, Geneva, 23, Switzerland \\
and $^{2}$Department of Physics, FM-15                 \\
University of Washington, Seattle, WA 98195, USA

\vspace*{1cm}

{\bf Abstract}
\end{center}

 We show that the low frequency, long wavelength dynamics of the phase
of the pair field for a BCS-type s-wave superconductor at T=0 is
equivalent to that of a time-dependent non-linear Schr\"odinger Lagrangian
(TDNLSL), when terms required by Galilean invariance are included. If the
modulus of the pair field is also allowed to vary,
the system is equivalent to
two coupled TDNLSL's.
\vspace*{0.5cm}

\noindent
PACS${\#}$s:74.20.-z, 67.50.Fi
\vspace*{1cm}
\begin{flushleft}
CERN-TH.7385/94\\
November 1994 (revised)

\end{flushleft}

\noindent
$^{\dag}$ Permanent address: Physics Department, Theoretical Physics\\
1 Keble Road, Oxford OX1 3NP, UK\\
$^{\#}$ Present address: Department of Theoretical Physics \\
                Ume\aa{\ }University, S-901 87, Ume\aa, SWEDEN \\
\vfill\eject

The classic Ginzburg-Landau (GL) theory [1] is very successful[2,3]
in describing a large class of static superconducting phenomena near the
critical temperature $T_{c}$, and its form was established by Gorkov [4]
shortly after the microscopic BCS theory [5]. Subsequently, a
number of attempts [2,3,6,7] were made to obtain a generalized GL
theory for time-dependent phenomena, and for temperatures well
below $T_{c}$, but a consensus has still not been reached on the
form of such a theory at T=0. In this paper we shall show that the
low frequency,long wavelength dynamics of the phase of the
pair field for a BCS-type s-wave superconductor
 at T=0 is equivalent to a time-dependent non-linear
Schr\"odinger Lagrangian (TDNLSL).
  At first sight, this result might seem almost obvious: after all, the
energy density in GL theory looks formally like that of a non-linear
Schr\"odinger theory so that it
seems natural to extend it to the corresponding
time-dependent theory as, indeed, Feynman assumed [8] in his
discussion of the dynamics of superconductors and of the Josephson effects.
Yet neither the earlier discussions[2,3,6], nor recent work based on the
effective action formalism of quantum field theory [7,9], appears to lead
to this conclusion. This is in contrast to the case of a Bose superfluid,
such as $^{4}$He, which is well described by a TDNLSL near T=0[10].
Indeed, there is considerable current interest in probing the relationship and
``crossover'' between BCS and Bose superfluidity[11]. Our result implies that
both are fundamentally the same, at least near T=0
 in the clean limit ; in particular, the
existence of the Magnus force for a vortex line in a superconductor
follows naturally. The last point is pertinent to the discussion of vortex
dynamics in superconductors within the effective theory formulation [12].

   Three of the present authors have, in fact, recently shown [13] that
the motion of the condensate is described by a non-linear Schr\"odinger
equation at T=0, using a density matrix approach
and the Born-Oppenheimer approximation. But this left open
the question how this could be reconciled with the earlier
work [2,3,6,7,9], which was generally based on field theory (or
Green function) techniques, and apparently led to a quite different result. The
solution of this problem is contained in the present paper,
and it is essentially very simple.
In [13] the further approximation was made that
the modulus of the energy gap (or pair field) is constant.
If this approximation is made in the field theory approach,
one can derive [9] from BCS theory an effective
action for the phase $ {\theta (x)}$ of the
pair field (i.e. the Goldstone mode), which one then expands up
to a certain order of derivatives. The resulting Lagrangian
 $L_{eff}(\theta)$ is the
same as that previously proposed on symmetry grounds [14], and also
corresponds precisely to the early results of Kemoklidze and
Pitaevskii [15], who started from Gorkov's equations [5]. We shall show
that the dynamics of $\theta(x)$ as given by $L_{eff}(\theta)$ can
be re-written in terms of a TDNLSL, which is equivalent to a
particular case of the general non-linear Schr\"odinger theory
derived in [13] under the same approximation.

We also extend this to include variations in the modulus of the pair
fields, and show that the dynamics is then that of two
 coupled TDNLSL's. The thread that unites all these approaches is
ultimately Galilean invariance. Since the microscopic starting point is always
Galilean invariant, one expects any theory
based on an effective action to preserve
this symmetry, a point emphasized in Ref. 15, and the
Schr\"odinger Lagrangian is the simplest such available.

We begin with the BCS Lagrangian
for s-wave pairing and in the absence of external fields:
 \be
L = \sum_{\sigma}\psi^{\ast}_{\sigma}(x)\left( i\partial_{t} +
\frac{\nabla^{2}}{2m} + \mu \right)
\psi_{\sigma}(x) + g\psi^{\ast}_{\uparrow}(x)
\psi^{\ast}_{\downarrow}(x)\psi_{\downarrow}(x)\psi_{\uparrow}(x)
\ee
 where $\psi_{\sigma}$ describes electrons with spin $\sigma=(
\uparrow, \downarrow )$, $\mu=k_{F}^{2}/2m$ is the Fermi energy in the
normal state, and $x=({\bf x}, t)$. Introducing the auxiliary(pair) fields
$\Delta(x)$ and $\Delta^{\ast}(x)$, and integrating out the electron
fields, one obtains the effective action
\be
S[\Delta, \Delta^{*}]=-iTr \ln G^{-1} -
\frac{1}{g}\int d^{4}x|\Delta(x)|^{2}
\ee
where the Nambu Green function satisfies
\be
\left( \begin{array}{cc}
O_{1} & \Delta(x) \\
\Delta^{*}(x) & O_{2}
\end{array} \right)
G(x, x') = \delta(x-x')
\ee
with $O_{1} = i\partial_{t} + \nabla^{2}/2m + \mu$,
$O_{2} = i\partial_{t} - \nabla^{2}/2m -\mu $, and Tr includes interval
and  space-time indices. To obtain from (2) an
effective Lagrangian in terms of
the degrees of freedom represented by $\Delta$, a possible
procedure [7] is to set $\Delta(x) = \Delta_{0} + \Delta'(x)$ where
$\Delta_{0}$ is the position of the minimum of $S$ for space-time
independent $\Delta$, and where $\Delta'$ is assumed to be slowly varying
in both space and time. One then expands $Tr \ln G^{-1}$ in powers
of derivatives of $\Delta'$. There are, however, two (related) objections
to this. First, we are dealing with the spontaneous breaking of
a local $U(1)$ phase invariance, which implies that at a temperature
far from the transition temperature, the most important
degree of freedom is the phase of $\Delta$, which is the relevant
Goldstone field. It is this field, rather than the real and/or
imaginary parts of $\Delta$, which should carry the low
frequency and long wavelength dynamics.
Secondly, the ansatz $\Delta(x)
=\Delta_{0} + \Delta'(x)$ violates the Galilean invariance possessed by (1),
which implies [15] that
\be
\Delta({\bf r}-{\bf v}t,t)\;
\exp(2im{\bf v}\cdot{\bf r} -imv^{2}t)
\ee
should satisfy the same equation of motion as $\Delta({\bf r},t)$. We
shall return to the question of Galilean invariance below.

  We therefore set
\be
   \Delta (x) = e^{i\theta (x)}|\Delta(x)|
\ee
and
$|\Delta(x)| = |\Delta_{0}| + \delta|\Delta(x)|$,
where we are interested in the low
frequency and long wavelength fluctuations of $\theta(x)$
and $\delta|\Delta(x)|$.
However, although $\delta|\Delta(x)|/|\Delta_{0}|$ is expected to be small,
and a simple
expansion of the sort mentioned above for $Tr \ln G^{-1}$ could easily
be set up in terms of derivatives of $\delta|\Delta(x)|$
if $\theta(x)$ were zero,
it is crucial to recognize that $\theta(x)$ is not small in general,
so that the phase factor in
(5) cannot be expanded, but must be treated as a whole. This prevents
a straightforward expansion of $Tr \ln G^{-1}$ when (5) is substituted
into (2). Fortunately, this difficulty can be easily circumvented [9,16].
Defining $U(x)=\exp(i\theta(x)\tau_{3}/2)$, we can write
\be
   Tr \ln G^{-1}=Tr \ln G^{-1}UU^{-1} = Tr \ln U\tilde{G}^{-1}U^{-1}
    = Tr \ln \tilde{G}^{-1}
\ee
where
\[
   \tilde{G}^{-1}=G^{-1}_{0} ( 1 -G_{0}\Sigma)
\]
\be
   G_{0}^{-1} = \left( \begin{array}{cc}
               O_{1} & |\Delta_{0}| \\
              |\Delta_{0} | & O_{2}
                       \end{array} \right )
\ee
and
\be
   \Sigma =- i \nabla^{2}\theta /4m - i\nabla\theta\cdot\nabla /2m +
    (\dot{\theta}/2+(\nabla\theta)^{2}/8m)\tau_{3}
     - \delta|\Delta| \;  \tau_{1}  .
\ee
Minimizing (2) with $\theta=\delta|\Delta|=0$
yields the usual gap equation for
$|\Delta_{0}|$. The dynamics of $\theta$ and $|\Delta |$
is contained
in
\be
   S_{eff}[\theta, \delta|\Delta|]
     = i Tr \sum_{n=1}^{\infty}\frac{1}{n}(G_{0}\Sigma)^{n}
       -\frac{1}{g}\int |\Delta|^{2}d^{4}x \; ,
\ee
where we note that $\Sigma$ contains just $\delta|\Delta(x)|$
and derivatives of
$\theta(x)$, in terms of which (assumed small) quantities a useful
expansion can be conducted, following standard techniques [17].

  We now concentrate on $\theta(x)$, and set $\delta|\Delta|=0$
for the time being. Carrying out the calculation to the indicated order
in derivatives we obtain
\be
L_{eff}(\theta) = - \rho_{0}(\dot{\theta}/2 + (\nabla\theta)^{2}/8m)
                  + N(0)(\dot{\theta}/2 + (\nabla\theta)^{2}/8m )^{2}
\ee
where $\rho_{0}=k_{F}^{3}/3\pi^{2}$ is the electron density at T=0,
$N(0)$ is the density of states ( for one spin projection) at the
Fermi surface and we have adopted a convenient normalization;
note that $N(0)=\rho_{0}/2mv_{a}^{2}$ where $v_{a}=v_{F}/\sqrt{3}$
is the velocity of propagation of the Bogoliubov-Anderson mode.
Eqn. (10) is the same as Eqn. (4) in Ref.9 if $\theta$ is
replaced by $2\phi$ (see also Eqn.(2.6) of ref.14).
The equation of motion for $\theta$ which follows from (10) is
\be
\frac{\partial\rho}{\partial t} + \nabla\cdot {\bf j}=0
\ee
where
\be
\rho=\rho_{0}-N(0)(\dot{\theta} + (\nabla\theta)^{2}/4m)
\ee
and
\be
{\bf j}=\rho \nabla\theta /2m  .
\ee
Equations (11)-(13) are, in fact, precisely those obtained (to this
order in derivatives)
by putting
$\delta|\Delta|=0$ in Eqns.(21)-(23) of Ref.15. We now show how
the dynamics contained in (11)-(13) can be reinterpreted in terms of
a TDNLSL.

We  first remark that the forms of (11) and (13) strongly suggest
that the quantities $\rho$ and ${\bf j} $ have the physical
significance of a number density and of a number current density,
respectively. In fact, simple linear response theory (assuming,
as always, a derivative expansion) gives [18]
\be
\delta\rho \approx -N(0)\dot{\theta},\;
{\bf j}\approx \rho_{0}\nabla\theta /2m
\ee
where $\delta\rho$ is the departure of the density from the
equilibrium value $\rho_{0}$, and ${\bf j}$ is the number
density current. We now consider the implications of
Galilean invariance. For densities $\rho$ and ${\bf j}$
obeying (11), Galilean symmetry requires
\be
\rho'({\bf r}',t')=\rho({\bf r},t),\;
{\bf j}'({\bf r}',t')={\bf j}({\bf r},t)-{\bf v}\rho({\bf r},t)
\ee
where ${\bf r}'={\bf r}-{\bf v}t$,$t'=t$; that is, $\rho$ and
${\bf j}$ transform as $t$ and ${\bf r}$ respectively. From the
discussion around Eqn.(4) above, the behaviour of $\theta$
under Galilean transformations is given by
\be
\theta'({\bf r}',t')=\theta({\bf r},t)+mv^{2}t-2m{\bf v}\cdot
{\bf r} ,
\ee
from which it follows that $\delta\rho$ as given by (14)
is not invariant, and that ${\bf j}$ transforms
incorrectly. However, although $\dot{\theta}$ is not a Galilean
invariant, the combination $\dot{\theta} + (\nabla\theta)^{2}/4m$
is.(This is actually the justification for singling out the
terms given in (10) from the total $n=2$ contribution in (9).)
 Thus the requirement that $\delta\rho$ be Galilean invariant leads
precisely to the expression (12) for $\rho - \rho_{0}$, which can now
be identified as $\delta\rho$. Similarly, if we replace $\rho_{0}$
in the expression (14) for ${\bf j}$ by $\rho$, we find that ${\bf j}$
 transforms correctly and the expression for ${\bf j}$ is that in (13).

We are therefore led to seek a theory involving two fields
$\rho$ and $\theta$, such that the equation of motion for
$\theta$ is (11) and that for $\rho$ is (12),
with $\rho$,
$\theta$ and
${\bf j}$ related by (13). Consider the TDNLSL
\be
L_{\psi} = i\psi^{\ast}\dot{\psi} - \frac{1}{4m}\nabla\psi^{*}
\cdot\nabla\psi - V
\ee
where the mass has been chosen to be $2m$, and $V$ will be assumed
to be a function of $|\psi |$ only. Let us set
\be
   \psi = \sqrt{\rho}\;\exp(i\theta) .
\ee
Then inserting (18) into (17) and discarding a total
derivative, $L_{\psi}$ becomes
\be
L_{\psi} = -\rho\dot{\theta} - \rho (\nabla\theta)^{2}/4m
-(\nabla\rho)^{2}/16m\rho - V(\rho) .
\ee
The equation of motion for $\theta$ is then (11), with ${\bf j}$
given by (13), while that for $\rho$ is
\be
\frac{dV}{d\rho} = -(\dot{\theta} + (\nabla\theta)^{2}/4m )
-(\nabla\rho)^{2}/16m\rho^{2} + \nabla^{2}\rho/8m\rho^{2} .
\ee
If we now choose
\be
V=(\rho-\rho_{0})^{2}/2N(0)
\ee
and solve (20) by expanding in derivatives, we recover
precisely (12) at the relevant order.

Thus the introduction of the auxiliary variable $\rho$, which can be
expressed in terms of $\theta$ via its equation of motion, has allowed us
to rewrite the dynamics of the Goldstone field $\theta$, as given by
$L_{eff}(\theta)$, in terms of the TDNLSL (17). The variable $\rho$
is interpreted physically as the number density. It must be
stressed, however, that the wavefunction $\psi$ introduced in this
way (see (18)) is quite distinct from the pair field $\Delta$,
despite the fact that they have the same phase $\theta$. In our
development thus far, the modulus $|\Delta|$ has been held fixed,
whereas $\rho$ varies; there is no simple relation between
$|\Delta|$ and $\sqrt{\rho}$.

The dynamical theory described by (17) is a special case of the
general time-dependent non-linear Schr\"{o}dinger theory for the
condensate wavefunction derived in [13], in which the form of the
potential $V$ was not explicitly calculated. Here, it has been
necessary to fix $V$ in order to carry out the elimination of
the variable $\rho$, to the required order in derivatives. (We remark
that in [13] $\psi$ is normalised to the density of Cooper pairs,
rather than, as here, to the electron density.)

Before discussing the modifications caused by the inclusion of
the field $\delta|\Delta(x)|$, we make one further comment on
the Lagrangian (10). A simple alternative route to (10) is to
start from a Lagrangian which describes just the Bogoliubov-
Anderson mode, viz.
\be
L_{a} = \frac{1}{2}\dot{\theta}^{2} - \frac{1}{2}v_{a}^{2}(\nabla\theta )^{2}.
\ee
Now (22) is clearly not invariant under (16).
But, as we have seen, the combination $\dot{\theta} +~(\nabla\theta)^{2}/4m$
is invariant. Hence if in (22) we replace $\dot{\theta}^{2}$ by
$(\dot{\theta} + (\nabla\theta)^{2}/4m)^{2}$ and $(\nabla\theta)^{2}$
by $4m\dot{\theta} + (\nabla\theta)^{2}$ we will have a
Galilean invariant Lagrangian; and the result of these replacements is just
proportional to $L_{eff}(\theta)$. Actually, the $(\nabla\theta)^{2}$
term in (22) is of course invariant by itself, up to constants and a
total derivative. Indeed, the term $\dot{\theta}$ introduced above,
and present in $L_{eff}(\theta)$, is also a total derivative and does
not affect the equations of motion. Nevertheless it is important
physically, as it ensures that the density $\rho$ has the equilibrium
value $\rho_{0}$ [14].

  In view of its relative unfamiliarity, it maybe worth
noting that $L_{eff}(\theta)$ (or equivalently $L_{\psi}$) embodies the
usual phenomenology of superfluid dynamics at T=0 (see, for example
[19,20]). We identify $\nabla\theta/2m$ with the superfluid
velocity ${\bf v}_{s}$, and multiply $\rho$ and ${\bf j}$ of (12), (13)
by $m$ to convert them to mass density and flux,
$\rho_{m}$ and ${\bf j}_{m}$.
Eq. (11) is then the law of mass conservation, following from the fact
that $L_{eff}$ does not depend explicitly on $\theta$.
Eq.(12) is equivalent [15] to Bernoulli's equation,
if we make use of
$\delta\rho\approx 2N(0)\delta\mu$ and $\delta p\approx
\rho_{0}\delta\mu$. Since $L_{eff}$
does not depend explicitly on $t$, we have the energy conservation relation
\be
   \frac{\partial E}{\partial t} + \nabla\cdot{\bf Q} = 0
\ee
where using the canonical definitions (with suitable normalization),
 one finds
\be
   E\approx\frac{1}{2}\rho_{m}v_{s}^{2}, \;
   {\bf Q}={\bf j}_{m} \left( \frac{1}{2} {\bf v}_{s}^{2}
                              + \delta\mu \right),
\ee
and we have dropped a quantity of order $\delta\rho\;\delta\mu$ in
$E$.
Finally, since $L_{eff}$ is translation invariant we have the momentum
conservation relation
\be
\frac{\partial {\bf j}_{m} }{\partial t} + \nabla\cdot\Pi =0
\ee
where the momentum flux density tensor is
\be
\Pi_{ij} = \rho_{m}v_{si}v_{sj} + \delta p \; \delta_{ij}.
\ee
Eq.(25) is equivalent to Euler's equation. In Ref.14,
 the proportionality between
the momentum density and the number current ${\bf j}$ (defined by
$\partial L/ \partial (\nabla\theta)$), which is included in (25),
was taken as a constraint on possible Lagrangians $L$. If $L$ is a
function solely of the Galilean scalar $g=\dot{\theta}+(\nabla\theta)^{2}/4m$,
then
\be
   \frac{\partial L}{\partial\dot{\theta}} = \frac{\partial L}{\partial g}, \;
   \frac{\partial L}{\partial(\nabla\theta)}=\frac{\partial L}{\partial g}
   \frac{\nabla\theta}{2m} .
\ee
Since $\theta$ is a phase variable, we can interpret $\partial L/\partial
\dot{\theta}$ and $\partial L/\partial(\nabla\theta)$ as being proportional
to a conserved number density $\rho$ and number current density ${\bf j}$
respectively, so that (27) becomes just (13).
 The momentum density is then
automatically proportional to ${\bf j}$.
 Once again, Galilean invariance is the
essential principle.

 We now turn to the inclusion of the field $\delta|\Delta(x)|$.
$L_{eff}(\theta,\delta|\Delta|)$ can be extracted from (9),
up to a given order in derivatives,
but calculations rapidly become laborious. For our present purpose, we will
simply use the result of Ref.15 which, using the normalization of
(10), gives

\[
  L_{eff}(\theta, \epsilon) = -\rho_{0}\left[ \dot{\theta}/2 +
   \frac{(\nabla\theta)^{2}}{8m} + \frac{(\nabla\epsilon )^{2} }{8m}\right]
   +N(0)\left[ \dot{\theta}/2 + \frac{(\nabla\theta)^{2} }{8m}
+ \frac{(\nabla\epsilon )^{2} }{8m}\right]^{2}
\]
\be
  +N(0)\left[ \dot{\epsilon}/2 +
 \frac{\nabla\epsilon\cdot\nabla\theta}{4m}\right]^{2}
  -3N(0)|\Delta_{0}|^{2}\epsilon^{2}
\ee
where $\epsilon(x) \equiv \delta|\Delta(x)|/(\sqrt{3} |\Delta_{0}|)$
and we have retained corresponding
terms in $\epsilon$ and $\theta$. The quadratic terms in $\epsilon$ yield the
amplitude collective mode with a finite gap found in Ref.18
(and are also in agreement with the
result of Ref.7); we have omitted higher powers of $\epsilon$. The term in
$\dot{\epsilon}/2$ is made Galilean invariant by the addition of
$\nabla\epsilon \cdot \nabla\theta/4m$, since $\epsilon(x)$ is a scalar.
  We now find that in order to rewrite (28) in ``Schr\"{o}dinger'' form
we have to introduce a second auxiliary density feld, which we
call $\rho_{\epsilon}$, in addition to the earlier density
$\rho$, which now becomes $\rho_{\theta}$. Thus four fields $(
\theta,\epsilon,\rho_{\theta}$ and $\rho_{\epsilon})$ are
required, and (28) is actually equivalent to two coupled
Schr\"{o}dinger equations. That
is, the equations of motion for $\theta$ and $\epsilon$ which follow
from (28) are identical to those arising from
\[
   L_{\psi_{1},\psi_{2}} =  i\psi_{1}^{\ast}\dot{\psi_{1}}
   - \frac{1}{4m}\nabla\psi_{1}^{*}\cdot\nabla\psi_{1}
  -(|\psi_{1}|^{2}-\rho_{0}/2)/N(0)
\]
\be
   + i\psi_{2}^{\ast}\dot{\psi_{2}}
 - \frac{1}{4m}\nabla\psi_{2}^{*}\cdot\nabla\psi_{2}
- (|\psi_{2}|^{2}-\rho_{0}/2)/N(0)
 +\frac{3}{2}N(0)|\Delta_{0}|^{2}[Im \ln(\psi_{1}/\psi_{2})]^{2}
\ee
where
\be
   \psi_{1} = \sqrt{(\rho_{\theta}+\rho_{\epsilon})/2}\;
                     \exp(i(\theta+\epsilon)), \;
   \psi_{2} = \sqrt{(\rho_{\theta}-\rho_{\epsilon})/2} \;
                     \exp(i(\theta-\epsilon)).
\ee
For example, corresponding to (12), we have
\be
   \rho_{\theta}=\rho_{0} - N(0)\left(
                 \dot{\theta} + \frac{(\nabla\theta)^{2}}{4m}
      +\frac{(\nabla\epsilon)^{2}}{4m}\right) =
      \frac{\partial L_{eff}(\theta ,\epsilon)}{\partial \dot{\theta}}
\ee
and
\be
   \rho_{\epsilon} = - N(0)\left( \dot{\epsilon} +
     \frac{\nabla\epsilon\cdot\nabla\theta}{2m}\right) =
     \frac{\partial L_{eff}(\theta ,\epsilon)}{\partial \dot{\epsilon}}.
\ee
Eqn.(29) represents a system of two TDNLSL's coupled via the ``mass'' term
in (28). Expressions for all the conserved quantities can be found as
before, and will include quantum corrections to the semiclassical results
of (23)-(26).

The inclusion of electromagnetism in the above formalism
is straightforward.
Consider the formulation in terms of $L_{eff}(\theta, \epsilon)$.
Since $\theta$
is the phase of a field with charge $-2e$, gauge invariance implies that
$\dot{\theta}$ and $\nabla\theta$ must appear in the combinations
$\dot{\theta}-2e A_{0}$ and $\nabla\theta + 2e{\bf A}$ ($e > 0$),
where $A_{0}$ and ${\bf A}$ are the electromagnetic potentials. The
field $\epsilon$, on the other hand, is electromagnetically neutral. The
leading order electromagnetic charge and current densities are obtained by
multiplying $\delta\rho$ and ${\bf j}$ in (22) by $-e$ and
making the above replacements for $\dot{\theta}$ and $\nabla\theta$. One then
obtains the usual results [7]. In terms of the Schr\"odinger formulation, one
simply makes the expected minimal coupling substitutions:
$i\partial_{t}\rightarrow i\partial_{t}+2eA_{0}$ and
$-i\nabla \rightarrow -i\nabla +2e{\bf A}$ in (14) or (29) (note from
(30)  that both $\psi_{1}$ and $\psi_{2}$ have charge $-2e$).

  When the above analysis is extended to higher order derivative terms,
it is clear on dimensional grounds that some characteristic scale must enter.
In fact, such higher terms enter in the form $\partial_{t}/|\Delta_{0}|$ and
$v_{F}\nabla/|\Delta_{0}|\sim\xi\nabla$ (see for example Eqn.(35) of [9]),
where $\xi$ is the coherence length. The basis of the expansion is
therefore the usual assumption [15] that the characteristic frequency
$\omega$,
and wavenumber $k$, of variations of $\Delta(x)$ satisfy $\omega\ll
|\Delta_{0}|$, $k\ll\xi^{-1}$. Indeed, (28)
already yields a static solution for $\epsilon$
which decays exponentially over
a characteristic distance $\xi/6$. Such a solution is of the type
expected far from a vortex core. Inclusion of appropriate higher
derivative terms should make possible some predictions about the
vortex core structure.

  The TDNLSL formulation provides, we believe, a simple and unifying
framework for the discussion of dynamical effects in BCS superconductors
at T=0. Results which have been known for many years [2,3,6,15], as well
as those obtained by quite different methods only
recently [9,14], are all seen to be in agreement with each other,
and with the TDNLSL formulation.

 {\it Note Added:} After completion of this work we received a copy
of a preprint by Michael Stone [21], in which a similar conclusion
is reached concerning the effective TDNLSL when $|\Delta|$ is constant.

{\bf Acknowledgements}:
This work was supported in part by the US National Science Foundation
under grant No. DMR-9220733. The work of I.J.R.A, while at the
University of Washington, was also supported by a grant from the
US Department of Energy to the nuclear theory group at UW, and by
the physics Physics Department, UW. I.J.R.A. is grateful to the members
of the UW Physics Department for their hospitality and support.

{\bf References}

1. V. I. Ginzburg and L. D. Landau, Zh. Eksp. Teor. Fiz. {\bf 20},
1064 (1950).

2. See, for example, N. R. Werthamer in {\it Superconductivity}, Vol.1,
edited by R.D. Parks, M. Dekker, New York, 1969.

3. M. Cyrot, Rep. Prog. Phys. {\bf 36}, 103(1973) gives a critical
review.

4. L. P. Gorkov, Zh. Eksp. Teor. {\bf 36}, 1918 (1959).

5. J. Bardeen, L. N. Cooper and J. R. Schrieffer, Phys. Rev. {\bf 108},
1175 (1957).

6. M. Dreschler and W. Zwerger, Ann. Phys. {\bf 1}, 15(1992).

7. H. T. C. Stoof, Phys. Rev. {\bf 47}, 7979 (1993).

8. R. P. Feynman, R. B. Leighton and M. Sands, {\it The
Feynman Lectures on Physics}, vol. III, Addison-Wesley, 1965.
See also R. P. Feynman, {\it Statistical Mechanics}, Reading, Massachusetts,
1972.

9. A. M. J. Schakel, Mod. Phys. Lett, {\bf B} 14, 927 (1990); see also
Ch. G. van Weert, in {\it Thermal Field Theories}, Proc. 2nd Workshop
on Thermal Field Theories and Their Applications, Tsukuba, Japan, July 1990,
edited by H. Ezawa, T. Arimitsu and Y. Hashimoto (North Holland, 1991).

10. E. P. Gross, Nuovo Cimento {\bf 20}, 454 (1961); L. P. Pitaevskii,
Sov. Phys. JETP {\bf 12}, 155 (1961).

11. See, for example, C. A. R. S\'a de Melo, Mohit Randeria and Jan R.
Engelbrecht, Phys. Rev. Lett. {\bf 71}, 3202 (1993).

12. A. T. Dorsey, Phys. Rev. {\bf B46}, 8376 (1992).

13. P. Ao, D. J. Thouless, and X.-M. Zhu, preprint.

14. M. Greiter, F. Wilczek and E. Witten, Mod. Phys. Lett. {\bf B3}, 903
(1989).

15. M. P. Kemoklidze and L. P. Pitaevskii, Zh. Eksp. Teor.  {\bf 50},
160(1966).

16. U. Eckern, G. Sch\"on and V. Ambegaokar, Phys. Rev. {\bf 30},
6419 (1984).

17. C. M. Fraser, Z. Phys. {\bf C 28}, 101 (1985); I. J. R. Aitchison
and C. M. Fraser, Phys. Rev. {\bf D 31}, 2605 (1985).

18. See, for example, E. Abrahams and T. Tsuneto, Phys. Rev. {\bf 152},
416(1966).

19. L. D. Landau and E. M. Lifshitz, {\it Fluid Mechanics},
Pergamon Press Inc, New York, 1959, Section 130.

20. M. J. Stephen, Phys. Rev. {\bf 139A}, 197 (1965);
O. Betbeder-Matibet and P. Nozieres, Ann. Phys. {\bf 51}, 392 (1969);
A. G. Aronov, Y. M. Gal'perin, V. L. Gurevish and V I. Kozub,
Adv. Phys. {\bf 30}, 539 (1981).

21.M.Stone,``Dynamics of $T=0$ BCS condensates'',University of
Illinois Physics Department preprint,ILL-(CM)-94-22,September 1994.

\end{document}